# A New Theory for Estimating Maximum Power from Wind Turbines: A Fundamental Newtonian Approach

[1,3]Ivan R. Kennedy, [2]Migdat Hodzic, [3]Angus N. Crossan, [1]Niranjan Acharige and [4]John Runcie

[1]School of Life and Environmental Sciences, Institute of Agriculture, University of Sydney
[2]Faculty of Information Technologies, The Dzemal Bijedic University of Mostar, Mostar 880000 Bosnia and Herzogovina
[3]Quick Test Technologies, c/- Institute of Agriculture, University of Sydney

**Abstract**
A novel method for calculating power output from wind turbines using Newtonian mechanics is proposed. This contrasts with current methods based on interception rates by aerofoils of kinetic energy to estimate power output, governed by the Betz limit of propeller theory. Radial action [$mr\omega r\delta\phi = @$, J.sec] generates torques from impulses from air molecules at differing radii on rotor surfaces, both windward and leeward. Dimensionally, torque is a rate of action [$(mr^2\omega\delta\phi)/\delta t$, $ML^2T^{-2}$, Nm]. Integration of the windward torque [$T_w$, Nm] is achieved numerically using inputs of rotor dimensions, the angle of incidence ($\theta$) of elastic wind impulse [$\delta Mv$] on the blade surface, chord and blade lengths and the tip-speed ratio with wind speed. The rate of leeward or back torque [$T_b$, Nm] in the plane of rotation is estimated from radial impulses from the blade's rotation on material particles, with magnitude varying with the square of the blade radius and its angular velocity. The net torque ($T_w - T_b$) from these rates of action and reaction is converted to power by its product with the angular velocity of the turbine rotors [$P = (T_w - T_b)\Omega$, Watts or J sec$^{-1}$], considered as an ideal Carnot cycle for wind turbines; its design should assist optimisation of the aerodynamic elements of turbine operation. A matter of concern must be predictions for a significant rate of heat production by wind turbines, represented partly by the magnitude of the leeward reaction torque but also by a greater release of heat downwind caused by a turbulent cascade in the wake of air flow following its impacts with the blades. Given the widespread occurrence of wind farms as sources of renewable energy and a need to minimise environmental impacts this new method should promote improved theory and practice regarding wind energy.

1. Introduction

Current models of wind turbine function use aerodynamic principles derived largely from aerofoil and propeller theories. The Rankine-Froude momentum and actuator disk models were developed in the 19th century with Betz and Glauert [1](1926; 1935) providing refinements related to wind turbine efficiency, including more recent developments [2](Sorensen, 2015). These models also include the axial motion of the air induced by rotors, a marked contrast to the radial action model that assumes turbine blades generate power while rotating into undisturbed air, taking no account of the down-wind wake. Similar action mechanics based on varying radial separation of molecules at different temperatures undergoing impulsive collisions was needed in our work on calculating the entropy of atmospheric gases [3](Kennedy et al., 2019) and in a revision of the Carnot cycle [4] (Kennedy and Hodzic, 2021a).

A detailed explanation of more recent blade element momentum theory (BEM) that accounts for angular momentum of the rotor need not be given here. However, in brief summary, BEM leads to an inexact expression for power output (*P*) according to the following equations (1) and (2) as a function of the cube of wind velocity ($v$), air density ($\rho$), the area swept by the rotor blades with diameter $D$ ($A_D$) and a specific axial induction factor ($a$) related to changes in angular momentum of the fluid.

$$P = [0.59v^3\rho A_D]/2 \qquad (1)$$

$$P = 2a(1-a)^2 v^3 \rho A_D \qquad (2)$$

This enables power extraction for a system that includes a rotating wake, which can be shown to give a maximum consistent with the Betz limit for power from kinetic energy of 0.593. Taken with other inefficiencies, power output of about 30% the theoretical maximum exerted form kinetic energy is found in practice.

Forces from air flow around an aerofoil dependent on angular dimensions of the blade are decomposed into lift and drag normal and tangential to the apparent wind speed. This enables forces rotating the turbine and those just bending the rotor to be calculated separately, taking the axial factor ($a$) of Equation (2) into account to estimate torques. However, we will point out flaws in this model, caused by mismatching the interception of wind momentum by the blades and a model of the inertial power of wind judged as inferior [4](Kennedy and Hodzic, 2021a).

## 2. Radial action theory

In the radial-action model for estimating wind power the details of blade aerodynamics are secondary and need only be considered later as refinements. Similar to Carnot's cycle for heat engines [4](Kennedy and Hodzic, 2021a) a radial action cycle for a wind turbine is considered as ideal, estimating the maximum possible motive power, without considering its production. Inefficiencies from friction or other causes will not be dealt with, to be considered in separate articles. Radial action mechanics applies to blades of any shape, while the fundamental differences in geometry and the torques generated by the windward ($T_w$) and leeward surface ($T_b$) are respected. Figure 1 models the torques generated on the blade surface areas, allowing the maximum power to be estimated as a function of windspeed, its angle of incidence and affecting actions and reactions in the blade surface material, controlled by the blade length ($L$ or $R$). chord width ($C$) and the tip-speed ratio [$L\Omega/v = \lambda$] of rotor rotation [$L\Omega$] compared to wind speed ($v$).

The main features of the impulsive action on the windward surface of the blade (Figs. 1, 2) follow.

(i) Impulses ($\delta mv = \delta mr\omega$) generated by material particles on elastic rotor surfaces as envisaged by Newton power the turbine's rotation at the hub, if free to do so. The air particles impact on trajectories imposing the inertia of the wind velocity on their far greater microscopic velocities, with mean free paths of the order of picometres. Action impulses [$\delta mvR$, J.sec] with $R$ the radial dimension to the hub, reflect the momentum of air trajectories from the surface. No detailed consideration need be given to the individual trajectories of the air molecules comprising wind, given that the transfer of momentum is collective. The rate of impulsive action [$T_w = \Sigma\delta mvR/\delta t$, J or N.m] provides the magnitude of the windward torque exerted.

(ii) The angle of incidence $\theta$ is also the angle of reflection from a flat surface Fig. 1), giving a total deviation angle of $2\theta$ for the momentum. The decreased forward velocity is the source of effective lift normal to the flat surface, determined by this angular deviation.

(iii) A turning moment is exerted within the blade material in the plane of freedom of rotation of the turbine, at an angle normal to the direction of wind incidence. The magnitude of the turning moment [$Mv\sin2\theta$] and its cause is illustrated in Figure 2.

(iv) This application of Newton's third law requires that the true wind direction be considered to estimate the reaction on the turbine blade. The apparent change of wind direction caused by the rotation of the rotors, critical in BEM theory, is irrelevant in the radial action model. The current blade momentum theory using aerofoil theory as an analogue assumes lift normal to the blade and drag in the same direction as the blade; taking account of axial air motion downwind is an unnecessary confusion of cause and effect for impulsive radial action.

(v) Only two vectors for force on the blades need be considered. The first is uniform with length along the blade from reflected windward impulses on the surface blades, giving the torque generating the rotation; the second vector is a reaction torque variable with blade radius $R$ from impacts by the rear of the advancing blades on air molecules, tangential to the direction of rotation (Fig. 1). At tip-speed ratios greater than one, the back torque involves impulses of greater magnitude than ever seen in aerofoils except when rapidly gaining altitude. As a result, for most of the blade, except perhaps that adjacent to the hub, no drag force is caused by turbulence on the downwind surface; drag has little or no analogue in wind turbines (Figures 1, 2). If the aerofoils of aircraft are considered as analogues for the blades of wind turbines, the aircraft should be rotating around its longitudinal axis as it is impelled forward by propellers or the thrust of jets. It is suggested this discrepancy could make BEM theory a flawed approach.

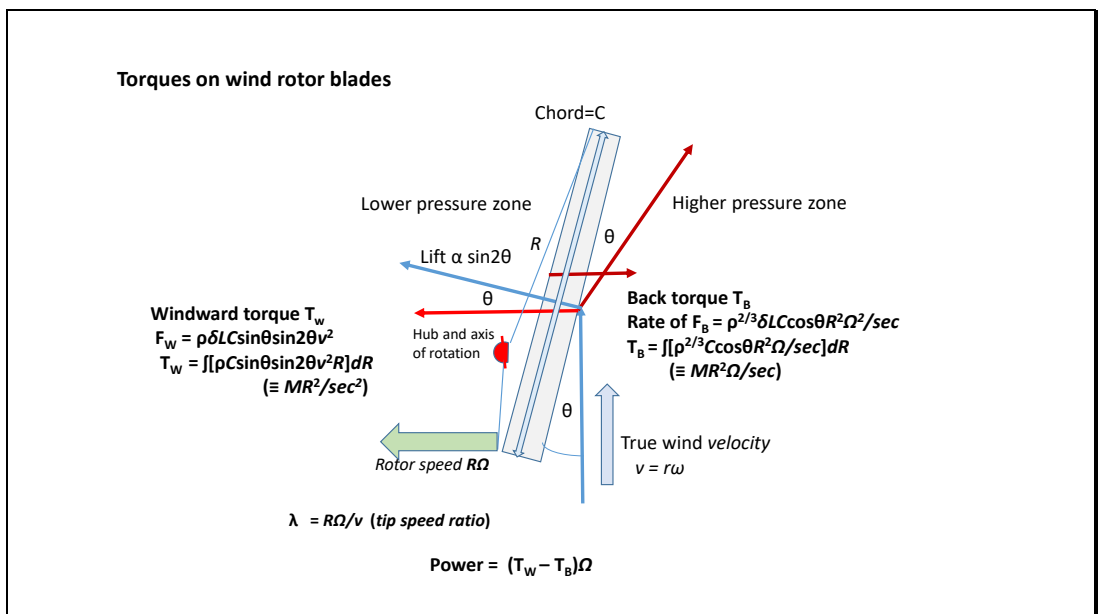

**Figure 1.** Windward ($T_w$) and leeward ($T_b$) back torques developed on a rectangular rotor blade. Equations were generated from trials using a numerical program using dimensions shown. The theory provides a theoretical maximum of power for a given angle of wind incidence, obtained by the difference between torques $T_w$ and $T_b$, occurring at about $\theta=60°$. Note the different dimensions employed for the density factor, explained in the text.

2.1 *Analysing of the rotor-turning moment from wind pressure*

Whatever paths the individual air molecules take in flow near the blade surface, dictated as laminar while the Reynolds number remains small, it is a fundamental principle of physics that the linear momentum at a given radial action is conserved, assuming perfect elasticity. By Newton's experimental law the impact of elastic bodies for oblique collisions on smooth surfaces will be reflected by the same angle for coefficients of restitution of 1.0. However, part of the moment exerted by wind particles may be extinguished if absorbed as thrust acting to push the wind tower and blades in a direction unable to rotate the rotor on its axis. For smooth surfaces there is no force parallel to the surface and the component of the particle velocity in the direction of motion is shown in Figures 1 and 2 as fully conserved with the angle of reflection equal to the angle of incidence.

For oblique impacts in the range 0-90 degrees, an asymmetric compression of blade material at the surface is generated as an oscillating function, dependent on the elasticity and density of the blade material. If the blade surface remains clean, this variation in stress as reactive pressure produces strains distorting the windward surface, varying the chemical potential in the compressed zones as a function of the angular deviation of the reactions. As a relationship between elastic stress and strain, this reaction can be described as a function of Young's modulus ($E$) for the rotor surface material, with surface stress σ or uniaxial force ($m\omega^2$) and strain ε equal to the distortion $\delta l/l$.

$$E = \sigma/\varepsilon \qquad (3)$$

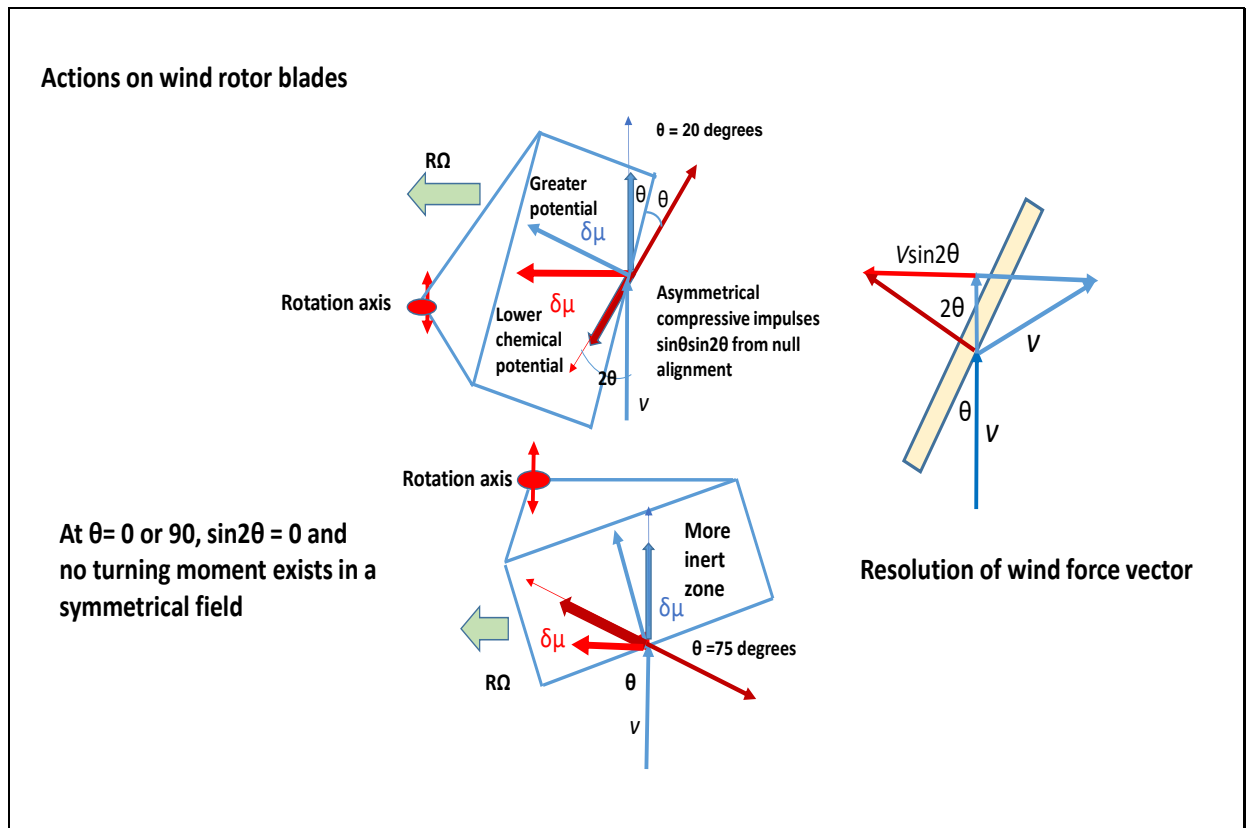

**Figure 2.** Generation of turning impulse ($V\sin2\theta$) from the surface normal to the wind direction is shown for two angles of incidence. Highly elastic action causes stresses and strain in reaction, intensifies chemical potential in rotor material, eased by rotation action exerted as a degree of freedom. For perfect elasticity of 1.0, the turning moment imparted to the rotor balances that of the reflected momentum in the opposite direction.

For the blade material distal to point of wind reflection, the physical reaction to the radial impulses is distributed in an arc of (90+θ) degrees while for the proximal reaction the arc for compression is (90-θ). The arc difference being 2θ, the turning moment per molecule on the rotor's axis is proposed to vary with $mv\sin2\theta$, thus balancing action and reaction. At 90 degrees or π/2 radians, this function becomes zero with any turning moment now symmetrical totally devoted to bending rather than turning the blade on its rotational axis at the hub. Obviously, this analysis can only be applied to compressions on the windward side of the rotor blades. The other factors determining the windward torque are density of air (ρ), the chord width ($C$) and sinθ, determining the volume and the mass of air impacting the blade per second. This represents the instantaneous magnitude of mass impacting per second for the area normal to the wind flux. When integrated with respect to radius ($R$) over the entire length of the blade ($L$) the cumulative torque ($T_w$) exerted at the hub is given in equation (4).

$$Tw = \int_0^L [\rho \delta LC \sin\theta \sin2\theta \, v^2 R]dR = \int_0^L [M\sin2\theta \, vR/sec]dR \qquad [ML^2T^{-2}, \text{J or N.m}] \qquad (4)$$

This equation comprises factors for the 3-dimensional density of air ($\rho$), the area of the blade at $R$ ($\delta LC\sin\theta$), the momentum per sec ($\rho\delta LC\sin\theta v = Mv$) made normal to the wind by $\sin\theta$, the extent of lateral reaction thus ($Mv\sin2\theta$). Numerically, the square of the wind velocity ($V^2$) is involved, once to estimate the mass of air impacting the blade per second and second to establish the magnitude of action impulse per second proportional to variation in action ($\delta mvR$) per molecule. This is considered as involving a rectangular blade in Figure 1 but different versions of the blade area at any radius can be estimated from variations in the chord width ($C$) as a function of the radius to the hub ($R$). The 3-dimensional density is regarded as a thermodynamic function, given that wind is a cooperative action with its inertia involving not just the kinetic energy of molecules striking the rotors on the windward surface but the vortical entropic energy and the resulting chemical potential, to be discussed in more detail in section 8.

2.2 *Leeward back torque of rotor blades*

It is said that a youthful Newton while constructing a windmill estimated wind force by the distance he could leap into the wind. This image is consistent with our model of the impact of the inertia of the blade on that of air. In contrast to the windward torque [$T_w$] proportional to radius, in which the wind factor $v^2$ applies uniformly over the rotor blade from hub to tip, the back torque ($T_b$) varies with the square of the radius, caused by the rate of impacts on the air behind the blade variable with $R$ during rotation. This variation is illustrated in Figure 2. Given that the speed of rotation [$R\Omega$, m sec$^{-1}$] determines both the mass of air impacted per second instantaneously as well at the radial momentum of these impacts, the specific action integral required is of radius squared [$R^2\Omega$]. An initial run of the model using $R^2\Omega^2$ as a factor was found to produce a power function rather than torque. This justified integrating the action per unit mass ($R^2\Omega$) instead of the energy per unit mass [$R^2\Omega^2$]. In effect, a variable inertial force along the blade is integrated with respect to velocity to provide the correct rate of impact with mass.

An important difference between windward and leeward impulses with air molecules lies in the irreversible nature of impacts from the blade on air molecules. While windward impulses may be considered as a balancing of forces from the wind on the blade, the leeward impacts generally exceed the wind speed except near the hub and cooperative resistance from air at the rear of the blade is much diminished and can be ignored. As a result, the density of air molecules is effectively exerted from 2-dimensional action impulses exerted as a series of minute slices of air, varying with radius. If the number density of molecules in a cubic metre is taken as proportional to n$^3$ then n$^2$ must represent the density of molecules the blade encounters as a continuous process. Taking the density as having a fractional exponent, the factor required should be $\rho^{2/3}$ or 1.145 rather than 1.225. By such a choice the correct physical dimensions to describe the rate of transfer of momentum from the blade to air molecules, integrated with respect to $R$, obtains the rate of impulsive action or reactive torque. The inertial mass ($MR$) impacted per second expresses the action function $MR^2\Omega$ rather than momentum $MR\Omega$, with decreasing orthogonality of impulses on shorter radii. To obtain the reverse torque, Equation (5) must be integrated

$$T_b = \int [\rho^{2/3} C\cos\theta R^2 \Omega / \sec]dR = \int [MR^2\Omega/\sec]dR \qquad [ML^2T^{-2}, \text{Nm}](\text{kg.m}^2 \text{per sec}^2) \qquad (5)$$

The surface area swept aside by the blade for each 1 metre segment of the length $L$ is $C\cos\theta$, given that the radius is varied to estimate torque for each decrease in the length of the blade. So the momentum generated in each second at each radius is equal to the volume swept aside per second [$C\cos\theta R\Omega$ x density $\rho$/sec = $MR\Omega$/sec]. Expressed as an action impulse depending on the radius, that gives action per sec or torque $\rho C\cos\theta R^2\Omega$/sec or $MR^2\Omega$/sec ($ML^2T^{-2}$). The longer the radius, the more orthogonal the impulse and the effectiveness of the action impact [$mrv$, J.sec].

Subtracted from the windward torque exerted on the front of the rotor blade to obtain the net torque on the rotor, then multiplied by the number of blades and by the angular frequency $\Omega$, the net power $P$ can be obtained.

Both torque equations can be derived with a constant value for any configuration of rotor operation and then integrated in a standard formula for $R$ and $R^2$ respectively [integrals of $L^2/2$ and $L^3/3$] for accurate outputs, assuming ideal conditions. Taking the derivative of factors such as angle of incidence, tip-speed ratio ($\lambda$) and rotor length ($L$) with respect to power allows optimisation of each of these factors. This should allow ease of control of these factors in wind turbine operation. Optimum tip speed ratios are usually in the range 3 to 10 with optimum length a function of wind speed.

Then turbine power ($P$) is estimated by the difference of windward and leeward torque multiplied by the angular frequency ($\Omega$).

$$P = [T_w - T_b]\Omega \qquad (6)$$

Equations (4) and (5) were determined as results by careful attention to physical dimensions ($ML^2/T^2$), confirmed by the variations in the torques observed in the numerical model, rather than from calculus of basic theory. A key specification was that the radial action wind turbine model should give good results for wide variations in the power outputs predicted, varying with rotor length and blade area from watts to megawatts.

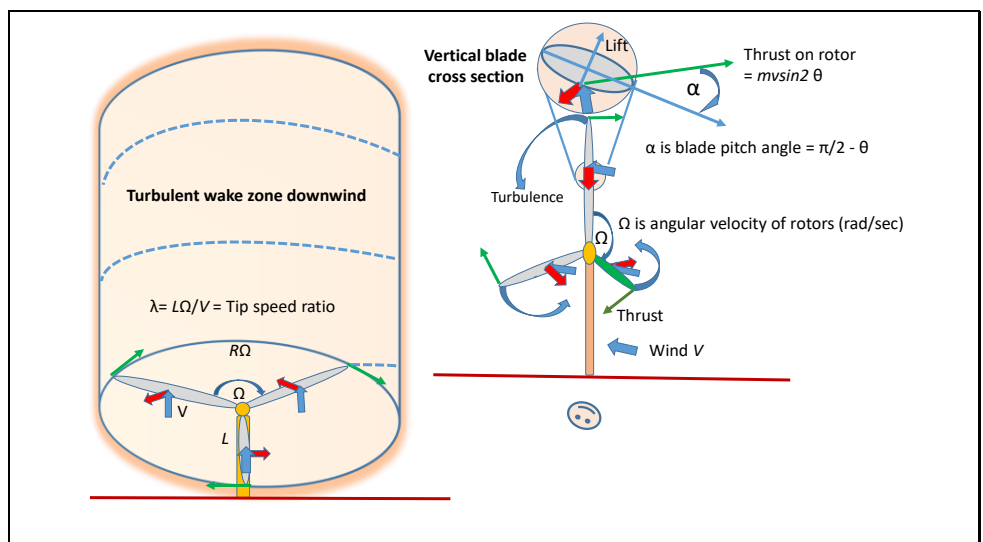

**Figure 3.** Plan and elevation of wind turbine showing clockwise motion caused by parallel pitching of all three blades ($\pi/2 - \theta$) to the plane of rotation with the thrust on the rotor normal to the direction of the wind stream, reflecting it accordingly ($\theta$). For a tip speed ratio ($\lambda$) equal to 12 for wind velocity ($V$) of 10 m/sec, the rotor tips travel 120 metres while the wind advances 10 metres.

Figure 3 shows a wind turbine with 40 metre blades similar to the General Electric (GE1.5) model, generating an average of some 1.5MW of power while rotating clockwise. Note that by contrast, Figures 1 and 2 would produce a counter-clockwise rotation observed from the windward direction. The blades reppresent about 3-4% of the area swept. While most of the air in the wind can pass between the blades unimpeded, some 20% of the air flow could be subjected to turbulent conditions. This suggests there is no requirement for a significant decrease in downstream windspeed, as is required using the BEM kinetic energy model because of the significant decrease in kinetic energy required. We argue in Section 5 that a lateral source of vortical potential energy [4](Kennedy and Hodzic, 2021a) sustains the kinetic energy, except in the turbulent volume (Fig, 3) where thermal energy will be released.

We emphasise that the main purpose of this article is limited to developing a testable hypothesis explaining the maximum power in an ideal wind turbine, assuming the output is linked reversibly to a work process such as electricity generation. This is similar to Sadi Carnot's proposal that the main purpose of his heat engine cycle was to describe the most efficient cycle. However, the environmental effect of turbulence will also be examined as it is a key consequence of the radial action hypothesis.

To allow testing of the radial action model for estimation of power outputs, approximate simulations of existing wind turbines were conducted using dimensions shown in Table 1. Blade lengths are advertised, but the maximum chords were estimates made from photographs. No account was taken of pitch values or twisting of the blade.

**Table 1. Unverified dimensions for simulating wind turbines**

| Example | Length (m) | Chord width (m) | Total blade area (m$^2$) |
| --- | --- | --- | --- |
| Vevor 500 mW | 0.520 +0.10 at hub | 0.023 -0.115 | 0.105 |
| GE 1.5MW | 38.75 + | 0.10-3.025 | 183.0 |
| Nordex N60 | 30 +1.0 | 0.25-3.25 | 120.0 |
| Nordex N131/3900 | 63+2.5 | 0.25-5.0 | 487.5 |
| Haliade-X 12MW | 100+ 8.5 stalk | 0.5-8.00 | 1290.0 |

### 3. Computer Model

The numerical computer scheme shown in Figure 4 employed inputs including wind speed ($v$), angle of wind incidence ($\theta$) and tip-speed relative to wind speed, defining a managed angular velocity [$\Omega$, radians sec$^{-1}$] for the turbines. The program was also designed to allow learning by experience. The complexity of current BEM models also suggested that a simpler means to determine wind power based on Newtonian physical principles might be needed. Our recent papers [4,5](Kennedy and Hodzic, 2021a,b) on action mechanics provided physical background to this work.

Elucidation of the governing equations followed analysis of test results using numerical methods. Optima in controlling factors were initially obtained by simple inspection of outputs. As shown in Figure 4, the program focussed on calculation the rate of action impulses [$\delta mvR/\delta t$] as torques, also calculating throughput of kinetic energy for comparison. Generated on a Windows platform with a capable Texas Instruments SR52 emulator, the program is available as Supplementary material has also been prepared using Mathematica with a Notebook suitable for blades with constant chord $C$ is also available in Supplementary material.

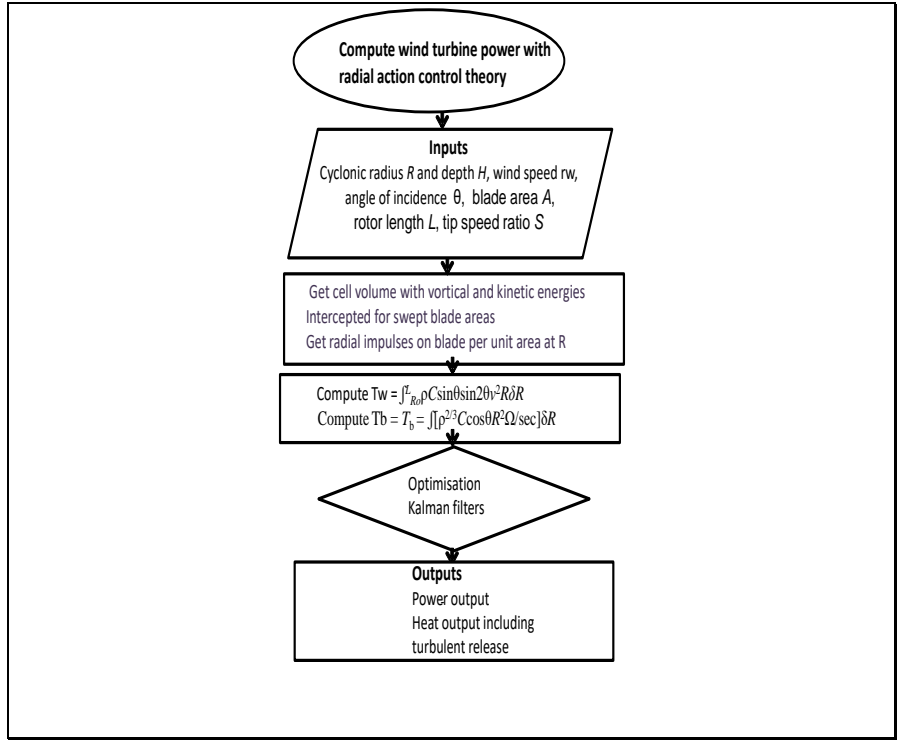

**Figure 3.** Flow scheme for computer program employed for numerical development of rotor theory. The coding is available as Supplementary material. Contact the corresponding author for more information.

## 4. Results and Discussion

Some representative results using the radial action model similar to those expected for a General Electric 1.5MW wind turbine of 83 m diameter are given in Table 2 and Figure 4. These show how the windward $T_w$ and leeward $T_b$ torques differ in character. The former shows a peak close to 55° diminishing towards 0° and 90°. By contrast, the leeward torque is maximal near zero degrees angle [θ] of incidence, decreasing slowly with the true angle of incidence to zero at 90°. Assuming a constant angular velocity, the potential power output shown in the figure has a similar form to the windward torque. The curves assume a given tip-speed ratio λ and rotational velocity. However, where there is no net torque ($T_w$-$T_b$) the rotor will not commence or will stall [7] at that angle of incidence. At a wind speed of 10 m per sec, the optimum value of λ was 8, but when wind speed is 20 m per sec, the optimum ratio was even greater than 20, a value challenging to strength of materials, since the tip speed would then be 400 metres per second, nearing the speed of sound in air.

**Table 2.** Selected data for Figure 4 showing variation in angle of incidence (θ) and tip-speed ratio (λ) for windspeed of 15 m per sec showing torques $T_W$ and $T_B$, turbine angular frequency (Ω) and power (P) outputs.

| V=15 | | | | | | | | | | | | | | | | | | |
|---|---|---|---|---|---|---|---|---|---|---|---|---|---|---|---|---|---|---|
| TSR=λ | 1 | 2 | 3 | 4 | 5 | 6 | 7 | 8 | 9 | 10 | 11 | 12 | 13 | 14 | 15 | 16 | 17 | 18 |
| θ=60 | | | | | | | | | | | | | | | | | | |
| Tw J | 6.12E+05 | 6.12E+05 | 6.12E+05 | 6.12E+05 | 6.12E+05 | 6.12E+05 | 6.12E+05 | 6.12E+05 | 6.12E+05 | 6.12E+05 | 6.12E+05 | 6.12E+05 | 6.12E+05 | 6.12E+05 | 6.12E+05 | 6.12E+05 | 6.12E+05 | 6.12E+05 |
| Tb J | 1.42E+04 | 2.83E+04 | 4.25E+04 | 5.67E+04 | 7.09E+04 | 8.50E+04 | 9.92E+04 | 1.13E+05 | 1.28E+05 | 1.42E+05 | 1.56E+05 | 1.70E+05 | 1.84E+05 | 1.98E+05 | 2.13E+05 | 2.26E+05 | 2.41E+05 | 2.55E+05 |
| Tw-Tb J | 5.98E+05 | 5.82E+05 | 5.69E+05 | 5.55E+05 | 5.41E+05 | 5.27E+05 | 5.13E+05 | 4.99E+05 | 4.84E+05 | 4.70E+05 | 4.56E+05 | 4.41E+05 | 4.28E+05 | 4.13E+05 | 3.99E+05 | 3.85E+05 | 3.71E+05 | 3.57E+05 |
| Ω | 3.87E-01 | 7.74E-01 | 1.16E+00 | 1.55E+00 | 1.94E+00 | 2.32E+00 | 2.71E+00 | 3.10E+00 | 3.48E+00 | 3.87E+00 | 4.26E+00 | 4.65E+00 | 5.03E+00 | 5.42E+00 | 5.81E+00 | 6.19E+00 | 6.58E+00 | 6.97E+00 |
| P Watts | 2.31E+05 | 4.51E+05 | 6.61E+05 | 8.60E+05 | 1.05E+06 | 1.22E+06 | 1.39E+06 | 1.54E+06 | 1.69E+06 | 1.82E+06 | 1.94E+06 | 2.05E+06 | 2.15E+06 | 2.24E+06 | 2.32E+06 | 2.39E+06 | 2.44E+06 | 2.49E+06 |
| λ=8 | | | | | | | | | | | | | | | | | | |
| θ | 5 | 10 | 15 | 20 | 25 | 30 | 35 | 40 | 45 | 50 | 55 | 60 | 65 | 70 | 75 | 80 | 85 | 90 |
| Tw J | 1.23E+04 | 4.85E+04 | 1.06E+05 | 1.79E+05 | 2.64E+05 | 3.53E+05 | 4.40E+05 | 5.15E+05 | 5.77E+05 | 6.16E+05 | 6.28E+05 | 6.12E+05 | 5.66E+05 | 4.93E+05 | 3.94E+05 | 2.75E+05 | 1.41E+05 | 0.00E+00 |
| Tb J | 2.26E+05 | 2.23E+05 | 2.19E+05 | 2.13E+05 | 2.06E+05 | 1.96E+05 | 1.86E+05 | 1.74E+05 | 3.08E+06 | 1.46E+05 | 1.30E+05 | 1.13E+05 | 9.58E+04 | 7.76E+04 | 5.87E+04 | 3.94E+04 | 1.98E+05 | 0.00E+00 |
| Tw-Tb J | -2.14E+05 | -1.75E+05 | -1.13E+05 | -3.37E+04 | 5.86E+04 | 1.57E+05 | 2.54E+05 | 3.42E+05 | 9.91E+06 | 4.70E+05 | 4.98E+05 | 4.99E+05 | 4.71E+05 | 4.15E+05 | 3.35E+05 | 2.35E+05 | 1.21E+05 | 0.00E+00 |
| Ω rad/sec | 3.10E+00 | 3.10E+00 | 3.10E+00 | 3.10E+00 | 3.10E+00 | 3.10E+00 | 3.10E+00 | 3.10E+00 | 3.10E+00 | 3.10E+00 | 3.10E+00 | 3.10E+00 | 3.10E+00 | 3.10E+00 | 3.10E+00 | 3.10E+00 | 3.10E+00 | 0.00E+00 |
| P Watts | -8.13E+06 | -5.42E+05 | -3.51E+05 | -1.04E+05 | 1.82E+05 | 4.86E+05 | 7.87E+05 | 1.06E+06 | 1.29E+06 | 1.45E+06 | 1.54E+06 | 1.54E+06 | 1.46E+06 | 1.29E+06 | 1.04E+06 | 7.29E+05 | 3.76E+05 | 0.00E+00 |

Complete Excel data set is included as supplementary data.

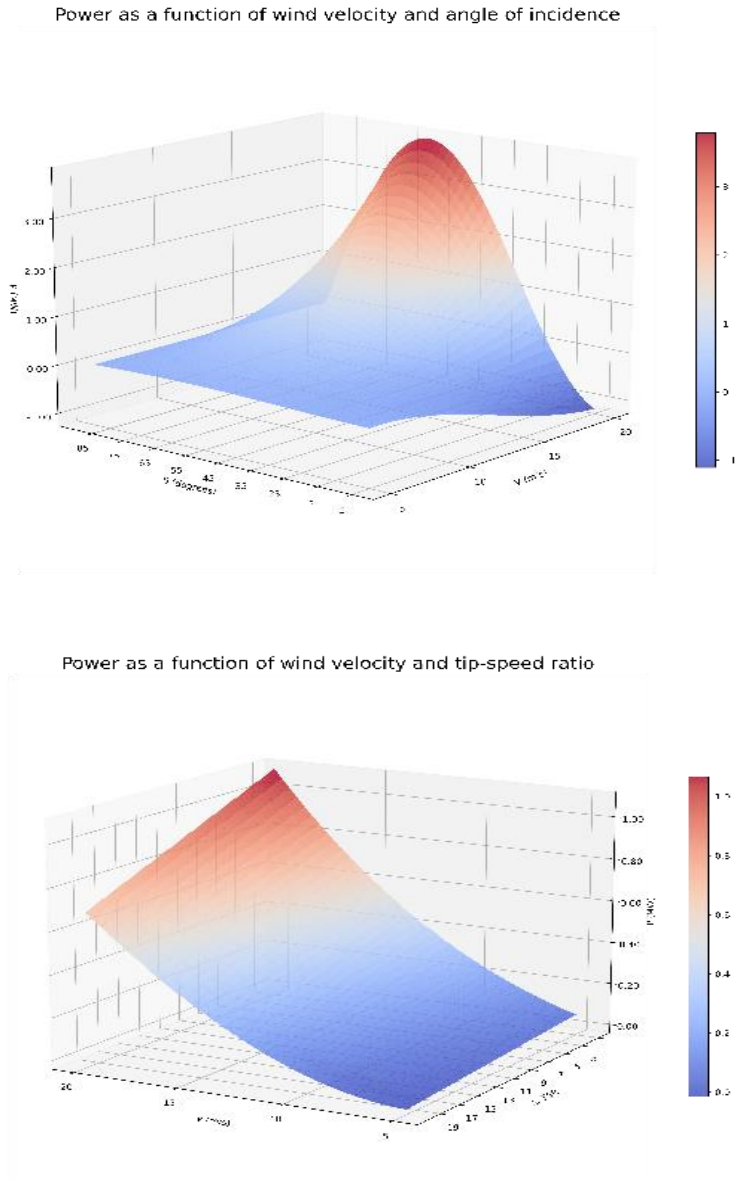

**Figure 4.** Power simulation of 1.5MW Turbine using the Turbine6 program given in supplementary material. A tip speed ratio of 8 was used for four wind speeds of 5, 10 15 and 20 m/sec used for the power diagrams, varying the angle of wind incidence θ (lower panel) and tip speed ratio (top panel) as shown.

For comparison, Table 3 gives results for associated wind kinetic energy with various wind speeds, showing data corrected for the Betz limit of 0.593 for the maximum power said to be extractable. Compared to the rate of kinetic energy passing into the area swept by the blades, the radial action prediction of power is always less, so if a mechanism to harvest were available, this is sufficient. However, when the kinetic energy in the wind actually impacting the blades is estimated, this is insufficient to explain the predicted power output. When the vortical entropic energy is estimated [5](Kennedy and Hodzic 2021b), This is much greater than the kinetic energy and always exceeds the power output by an order of magnitude.

**Table 3. Kinetic and vortical energy impacting a wind turbine similar to GE 1.5MW**

| Wind speed (m/sec) | Kinetic energy per sec 83 m diam. (Betz) (J) | Kinetic energy /blade-area/sec blades | Vortical pressure (J/m$^3$) (blade area xV | Vortical power (Watts, J/sec) estimated for blade area | Power estimated by radial action model (Watts) |
|---|---|---|---|---|---|
| | | | | | At $\lambda=9$, $\theta=55^\circ$ |
| 5.0 | 2.1670x10$^5$ | 8.4066x10$^3$ | 0.361069x10$^4$ | 0.33038x10$^6$ | 0.031168x10$^6$ |
| 10.0 | 1.7336x10$^6$ | 6.7253x10$^4$ | 1.47258x10$^3$ | 2.69482x10$^6$ | 0.40541x10$^6$ |
| 15.0 | 5.8509x10$^6$ | 2.2698x10$^5$ | 3.35055x10$^3$ | 9.19727x10$^6$ | 1.54381x10$^6$ |
| 20.0 | 1.3869x10$^7$ | 5.3802x10$^5$ | 6.00353x10$^3$ | 21.9729x10$^6$ | 3.86798x10$^6$ |

4.1 *Turbines of similar dimensions to commercial turbines*

A requirement for the radial action model was that it should be fully scalable with size. Figure 5 shows power results calculated for a wind turbine fitted with 52 cm blades (Table 1), similar to the Vevor commercial model. Vevor claim power of the order shown in Figure 5. These results were obtained for triangular blades of average chord width of 6.50 cm, tapering from 2.30 cm at the tip 62.0 cm from the hub to 11.50 cm at the base, supported on a 10 cm stalk to the hub.

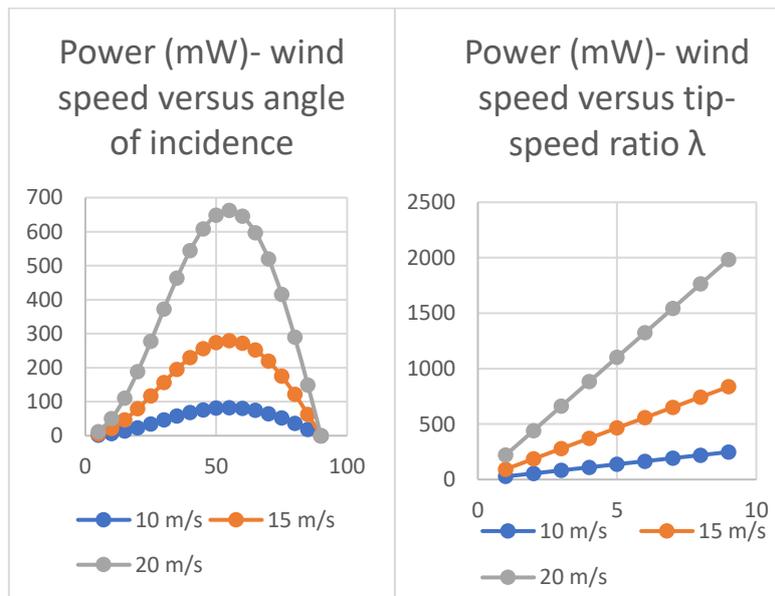

**Figure 5.** Radial action power output in milli-watts estimated for the Vevor 60 cm turbine showing optima for 55-60$^\circ$ wind incidence and increasing power with tip-speed ratio at all wind speeds.

Unlike the larger GE 1.5MW and Haliade-X 12MW turbines, the back torque ($T_b$) was negligible Table 4, reflecting the shorter radius and the $R^2$ factor involved. For the Vevor turbine the theoretical tip speed ratio ($\lambda$) shown in Figure 5 above 4 is excessive, exceeding the limit of 800 rpm by the manufacturer's guarantee. The program advanced chord width by 0.1769 cm for each reduction in radius of 1.0 cm.

When data for the Vevor turbine were employed for a rating wind speed of 12 metre per sec with $\lambda$ of 3.5, a power output of 406 Watts was calculated, This result is in good agreement with the published rating. This result is consistent with accuracy of the radial action model, but it should be experimentally tested such as in wind tunnels.

4.2 *Power estimates for simulated turbines*

Using the data on blade lengths and estimated chords in Table 1 a set of power outputs, assuming a triangular blade without twist with values of angle of wind incidence (θ) and tip speed ratio (λ) as given in Table 4. No attempt has been made in these estimates to optimise aerodynamics, with the windward and leeward surfaces considered as flat and fully elastic. However, it is anticipated that attention to the aerodynamics would produce marginal effects on power output.

**Table 4. Simulated power outputs in turbines similar to those in production**

| Brand | Blade (m) & angle (θ) | TSR (λ) $L\Omega/V_w$ | Wind m/sec | Chord C metres | Wind torque (Tw) MNm | Back torque (Tb) MNm | Power MWatts |
|---|---|---|---|---|---|---|---|
| Vevor 400 | 0.60, 60 | 3 | 15 | 0.112 | 0.00000363 | 0.000000031 | 0.00027 |
| GE 1.5MW | 38.75, 60 | 8 | 15 | 3.025 | 0.61191 | 0.11339 | 1.54380 |
| Nordex N60 | 30, 60 | 8 | 15 | 3.250 | 0.41318 | 0.07557 | 1.35043 |
| Nordex N131 | 80, 60 | 9 | 15 | 3.500 | 3.46481 | 0.76443 | 4.55715 |
| GE Haliade-X | 107, 60 | 10 | 15 | 8.000 | 11.41320 | 2.72240 | 13.0362 |

Only in the case of the Vevor and Nordex N60 was performance data available to the authors, with 0.400 kW and 1.13 MW respectively at 15 m per sec wind speed claimed. Of several large turbines examined, the Nordex claims were found to be the most modest with claims of 1.3 and 3.9 MW available in advertising material; several of the larger turbines about to be commissioned for marine settings seemed too optimistic, particularly when rating was conducted with wind speed close to 10 metres per sec. We suggest that advertised successful field trials over 24 hours may have been conducted with wind speeds greater than the rating speed.

From radial action theory, it is predicted that some of the even larger turbines planned for ocean platforms will not achieve performance anticipated. Given that the back torque is a function of the third power of the blade length when integrated, whereas the windward torque is a function of the blade length squared, this decrease in performance with increasing blade length is expected. These values in were all calculated with chord length diminishing from the hub to a small fraction of the maximum width, reducing this negative effect. No claim is made that the estimates in Table 4 can be considered as accurate. Chord widths are secret and have been estimated from photographs. No account has been taken of the rate of twisting of the blades. Such reduction in the pitch near the tip selectively reduces the back torque.

4.2 *Optimising Power*

We can estimate optimum values for wind turbine operation using calculus, for given environmental conditions of wind speed (*v*) or air density (ρ) or pressure. We can simplify the torque difference equation as follows using simplified versions of the equations discussed above separating windward and leeward constants ($K_w$ and $K_b$) from variables requiring integration.

$$\begin{aligned}T_w - T_b &= \int \rho C \sin\theta \sin 2\theta v^2 R dR - \int \rho^{2/3} C \cos\theta R^2 \Omega/\text{sec}.dR \\ &= K_w v^2 \int R dR \text{ from 1 to } L - K_b \int R^2 \Omega dR \\ &= K_w v^2 . L^2/2 - K_b \Omega L^3/3 \end{aligned} \quad (8)$$

The net wind torque equals the back torque when this expression is zero, that is, when the equality in (7) holds.

$$\begin{aligned} K_w v^2 L^2/2 &= K_b \Omega L^3/3 \\ L\Omega &= 3 K_w v^2 / 2 K_b \end{aligned} \quad (9)$$

The difference in torques is maximal when the derivative of $L$ with respect to power $P$ is zero.

$$P = K_w v^2 L^2 \Omega/2 \ - \ K_b \Omega^2 L^3/3$$
$$dL/dP = 0 = K_w v^2 L \Omega \ - \ K_b \Omega^2 L^2 \qquad (10)$$

Thus, maximum power is obtained when the following holds.

$$K_w v^2 L \Omega = K_b \Omega^2 L^2$$
$$L\Omega = K_w v^2 / K_b$$
$$K_b = \rho C \cos\theta$$
$$L\Omega = K_w v^2 / K_b \qquad (11)$$

Different winds and rotor lengths require modulation to different frequencies.

### 4.3 *Reconciliation of radial action and blade element momentum models*

The theoretical success of this model calls for some comparison with the blade element momentum (BEM) model based on aerofoil and Bernoulli fluid motion equations. In Figure 6 aspects of the two different approaches are given on one diagram, highlighting some differences. The radial approach we introduce involves two classes of action, quantifying inertial impulses of momentum at each radius of the rotor blade surfaces, one windward and the other leeward. In the figure, the pitched blades are regarded as rotating normal to the true wind direction with the turbines facing the wind turning anticlockwise. Air in inertial motion between the rotors is unimpeded with the blades reflecting windward impulses at the true angle of incidence ($\theta$). Only a small proportion of the air stream will impact the blades depending on the proportional area of the circle with radius $L$. If stationary, obstruction by the blade must create a region of low pressure to its rear. Once a steady state of rotation $\Omega$ is reached, the blade still obstructs air flow as before but its rear surface impacts resisting air, deflecting it with the action impulses a function of the radial speed of rotation ($R\Omega$). For most of the blade except near the hub, the pressure at the rear of the rotating blade must be increased above that in the wind. However, there is no diminution of pressure exerted on the windward surface because fresh air continually occupies this space.

The windward impact pressure is independent of the radius unless the blade is pitched with length, varying by $\sin\theta$ affecting the volume of air impacting the blades per sec. While the intensity of impulses is constant for a blade surface with constant pitch, the rate of action or torque varies with radius, requiring integration for the turbine. By contrast, the impulses produced by the leeward surface of the blades vary with the radius squared, once for the surface area of air impacted normal to the rotation per sec and once more for the radial variation in momentum As action impulses.

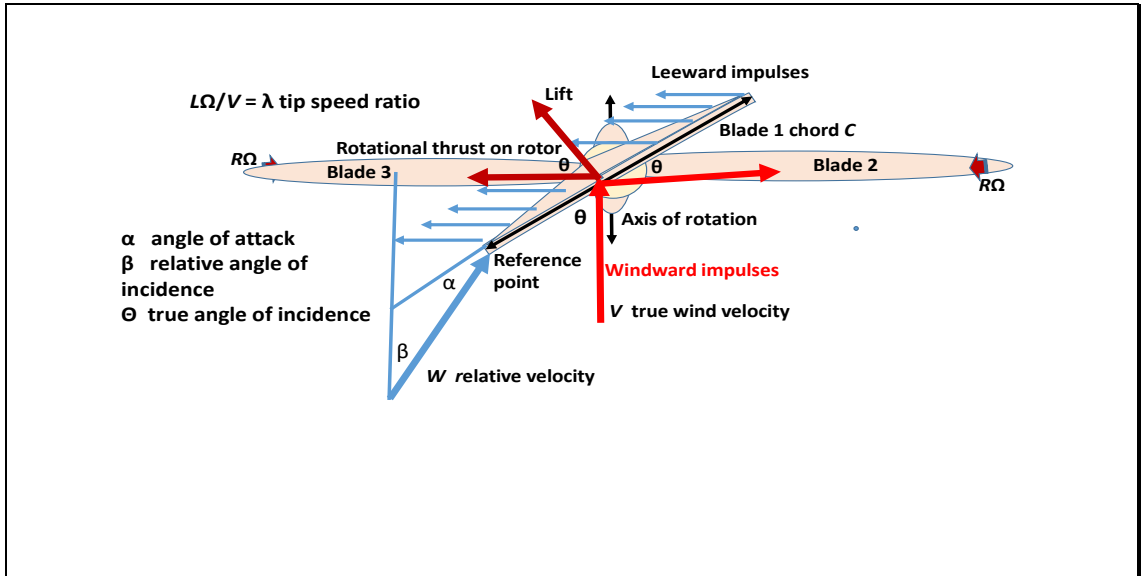

**Figure 6.** Comparison of blade element momentum (BEM) and radial action theory (RA). BEM subsumes the reverse or back torque ($T_B$) as an effect giving relative wind velocity ($W$). $T_w$ is derived by integrating the windward action impulses along the blade to the hub and $T_B$ integrates the backward push action impulses exerted by the blade.

There is no intention to describe BEM theory [6](Schubel et al, 2012) here in detail. However, in using a Bernoulli approach to aerofoil theory BEM considers lift normal to the blade and drag directed horizontally along the surface in the wind direction as the two main forces operating. To determine relative wind velocity the variable speed of an observation point on the rotating blade is taken. If near the hub, there is little change of direction (θ- β) but near the tip, the apparent angle of incidence diminishes. In Figure 6, the pitch angle to the chord line of the blade is 90-θ degrees, becoming zero when the true wind is normal to the blade, with no power to cause rotation. Alternatively, the pitch may be increased to 90º, when the back torque, while the blade is still spinning, will exceed the almost zero windward torque. Many of the performance studies on horizontal turbines have been performed with 'frozen' rotors [7](Sudhamshu et al. 2016), experiments conducted with variable speeds in wind tunnels enabling theoretical power to be calculated. However, BEM avoids consideration of back torque that impedes rotation.

4.4 *Wind turbine blade design, twist and other modulations in rotors*

In the BEM theory benefits justify twisting of the blade reducing the pitch towards zero degrees approaching the tip [6][Schubel and Crossley, 2012]. A twist is also justified in radial action diminishing the back torque proportional to $R^2$ nearer the tip whereas the windward torque is no more per *unit* area nearer the hub. This property is easily introduced into the radial action model by varying the relative angle of inclination towards the tip. Further aerofoil refinements commonly engineered into the rotors can easily be incorporated. These are considered to minimise frictional effects on turbines, making an independent contribution to efficiency of power output. Incorporating design features that are responsive to wind speed and other factors may optimise this process, that can be confirmed empirically.

This new understanding of power generation from radial action theory accepts the value of research on optimising blade design. Factors such as variation in thickness and twist of the chord pitch will still provide advantages in power output if correctly analysed. Schubel and Crossley [6] have provided a detailed review of the current state of the art of blade design, highlighting efficiency to be gained from design principles based on blade shape, aerofoil properties, optimal attack angles using relative wind speed and gravitational and inertial properties. With suitable corrections offered by the radial action approach to wind force, lift and thrust factors and generation of action most of this theory can be remodelled quite easily.

The Betz limit is considered to exert an effect on BEM theory but radial action challenges this. A new viewpoint on the nature of wind power in generating rates of action or torque is given in Section 5.

## 5. Wind speed and downwind turbulence

The calculation of wind power to the cube of wind speed shown in Equations 1 and 2 considers the rate at which kinetic energy of air flows through the circular profile area traced by the tips of the rotors. However, only a fraction of this air can be intercepted by the blades. Particularly in the larger modern turbines most of the air flow passes through unimpeded, given that the blades normally represent only some 3-4% of the area of the rotor circle. Thus, no more than 5% of the air volume is initially made turbulent, effectively tracing triplicate rotary spirals of turbulent air downwind, balancing the work done on the turbine rotors for transmission into the dynamo.

The leeward torque performs work normal to the air flowing into the cavity behind the blade, estimated as shown in Section 1. Additional release of radiant heat by turbulence is predicted to be a feature of the operation of wind turbines. We hypothesize a significant warming effect downwind that may also increase evaporation, caused by temperature increase and turbulent surface interaction with vegetation and soil surfaces. This prediction should be tested for quantification, to be included in productivity models or estimates of fire risk, as a matter of due diligence.

5.1 *Wind power in cyclonic air motion sustained as vortical action*

We have proposed [5](Kennedy and Hodzic, 2021a) that air in anticyclones and cyclones, subject to the inertial Coriolis effect, possesses a higher degree of freedom of action superior to the accepted vibrational, rotational and translational degrees of freedom. This increased source of entropy can be estimated from its vortical action (figure 7), capable of magnifying the heat capacity of air about two to three times, depending on the radius of action and the vortical frequency or wind speed ($v=R\omega$). By comparison, the kinetic energy of vortical motion has less than 5% of the same energy capacity. This is a testable hypothesis since it predicts that detectable thermal energy will be released as radiant heat from the cascade of turbulent conditions. Furthermore, colliding air masses must also generate radiant heat as laminar flow is replaced by turbulence. We [4] consider it is normal function of anticyclones that they should release radiant heat by friction with the surface, an important natural process transferring heat from the Hadley circulation of tropical air towards the poles. Too much interference with such natural energy flows could lead to intrusion of colder polar air; this may already be occurring in the polar vortices being experienced in both hemispheres.

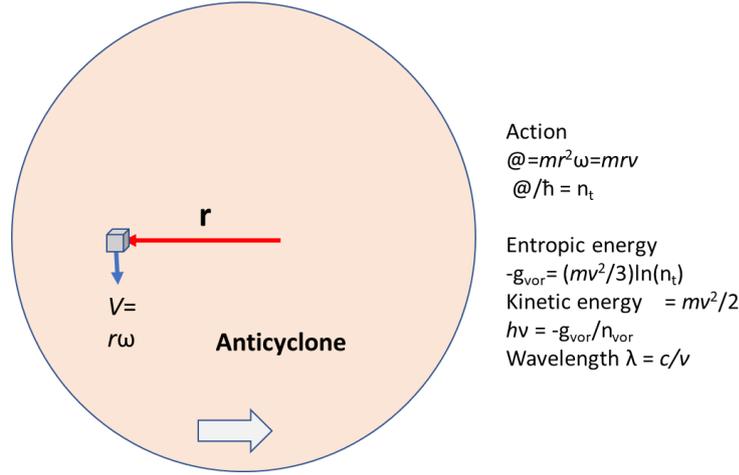

**Figure 7.** Vortical action and entropic energy in anticyclones as a source of wind power. The entropic or Gibbs field energy is proposed [4] as a function of action involved in the rotating airstream, as shown in the equations. The kinetic energy in wind is sustained by this vortical potential field.

The governing equations of fluid motion as formulated by Bernoulli, Laplace and others proposed no major reversible heat-work-heat processes, absorbing or releasing heat depending on whether air is expanding or being compressed. For streamlines as in a laminar wind flow, the Bernoulli equation (12) relates kinetic energy ($\rho v^2/2$), the static pressure energy $P$ ($\Sigma mv^2/3 = pV$) and gravitational potential energy, regarded overall in steady flow as constant.

$$\rho v^2/2 + P + \rho gh = K \tag{12}$$

The equation is also the basis of the theory that the pressure on the longer profile of an aerofoil will be lower, given that $\rho v^2/2 + P$ should be constant. The greater velocity required for air flow with a longer path requires that the pressure $P$ must fall, reducing the downward force. Despite widespread acceptance of this theory, clear evidence ifor confirmation is difficult to find.

The vortical entropic energy ($S_{vort}T$), hypothesised to be larger than the total heat content indicated by the entropy of air [3,4][Kennedy et al., 2019; Kennedy and Hodzic, 2021a), is additional to these forms of energy. Independently, the vortical component $S_{vort}T$ can be partly released in compressive or turbulent frictional processes. The same categories of energy transformation are also observed in the Carnot cycle, varying during the isothermal and the adiabatic stages. We have extended this hypothesis [6](Kennedy and Hodzic, 2021b) to show how this internal work that Clausius named the *ergal* amounts to a decrease in the Gibbs energy in Equation (13). To the extent that the vortical motion provides an additional degree of freedom for energy storage at a higher scale, another term needs to be added to the Bernoulli equation.

$$\rho v^2/2 + P + \rho gh + S_{vort}T = K + S_{vort}T \tag{13}$$

This can be thought of as a radial form of quantum state or potential energy, capable of being released in defined meteorological conditions.

Table 4 provides details of data for vortical entropic energy for a cubic metre of air as wind 1000 km from the centre of an anticyclone. The greater magnitude of the vortical component indicates that wind power is not so much a function of kinetic energy but more of the vortical energy exerting a torque. Calculation of vortical action (@$_{vor}$) and entropic energy is shown in equation (14), where $R$ is the radial distance to the centre of the cyclonic structure and $R\omega$ the steady windspeed at $R$.

$$@_{vor} = mR^2\omega; \quad mR^2\omega/\hbar = n_{vor}; \quad \text{Vortical energy per molecule} = kT\ln n_{vor} \qquad (14)$$

The relative vortical action and quantum state ($n_{vor}$) are proportional to wind speed but the vortical energy and quantum field pressure (field energy/unit volume) are logarithmic functions of the action as quantum numbers. The mean quantum size is exceedingly small and decreases with wind speed, most of the work or quantum pressure driving the motion of anticyclones (or cyclones) being acquired at lower temperature and wind speed. Given that air at 288.15 K and 1 atm pressure contains $2.5294 \times 10^{25}$ molecules of air per cubic metre, the vortical field energy of air is 3.505 kJ per cubic metre at a wind speed of 15 metres per sec. For this calculation we assumed a mean mass of 29 daltons for air to estimate action.

**Table 5. Vortical energy properties for GE 1.5 MW wind turbine**

| Wind speed (m sec$^{-1}$) | Vortical action (@$_v$)/molecule [J.sec, $\times 10^{19}$] | Quantum number $n_{vor} \times 10^{-15}$ | Vortical energy/ molecule [$(mv^2/3)\ln(n_{vor})$, $\times 10^{23}$ J] | Vortical quantum [$h\nu$, J $\times 10^{38}$] | Vortical energy [J/m$^3$] | Kinetic energy [J/m$^3$] |
|---|---|---|---|---|---|---|
| 5.0 | 2.4215 | 2.29615 | 1.42747 | 0.61210 | 361.068 | 15.313 |
| 10.0 | 4.8430 | 4.59230 | 5.82180 | 1.26677 | 1,472.580 | 61.250 |
| 15.0 | 7.2645 | 6.88845 | 13.24632 | 1.92298 | 3,350.556 | 137.813 |
| 20.0 | 9.6860 | 9.18975 | 23.73477 | 2.58419 | 6,003.530 | 245.000 |

In our previous paper [3](Kennedy et al., 2019) we showed that surface air heated from absolute zero to 298.15 K needs 2.4 MJ per cubic metre, including its kinetic energy. This implies that the vortical energy of air in laminar flow at 1000 km from the centre of an active anticyclone contains about 0.15% more field energy as at its non-rotating centre. The data in Tables 4 and 5 of possible heat capacities in wind suggests an alternative explanation. This source would be the release of latent heat in wind of vortical entropic energy, a result of turbulence caused by turbines. The concept of vortical entropy was advanced by [4](Kennedy and Hodzic, 2021a) as a new class of potential risk to be considered in climate change. The kinetic energy of wind of 10 m per sec is only 61 J per cubic metre.

Should the laminar flow be impeded by surface roughness, causing turbulence, some of the vortical energy will be released, warming the surroundings. For example, if wind speed of 15 m per sec is effectively reduced to a speed of 5 m per sec, some 7.5 MW or vortical turbulent heat is predicted to be released from air impacting the blades, heating the surrounding molecules. Given a heat capacity of 1.225 kJ per cubic metre for air, this is sufficient to heat 6,122 cubic metres of air 1 degree Celsius. A windfarm 1 km wide generating 100 MW of power from 70 GE 1.5MW turbines is predicted to release 750 MW of heat, moving 750,000 cubic metres of air 10 metres downwind a second, raising air temperature 1 degree Celsius in a column 75 m high. This prediction can readily be tested.

In developing turbulence, the largest scale eddies nearer laminar flow are regarded as containing most of the kinetic energy, whereas smaller eddies are responsible for the viscous dissipation of turbulence kinetic energy. Kolmogorov [7](Frisch, 1995) hypothesized that the intermediate range of length scales could be statistically isotropic, and that a temporary form of equilibrium would depend on the rate at which kinetic energy is dissipated at the smaller scales. Dissipation is regarded as the frictional conversion of mechanical energy to thermal energy, effectively radiation, raising temperature. In vortical action theory, the kinetic energy is regarded as always complemented by the Gibbs field vortical entropic energy and the dissipation process loses kinetic energy as a result of the loss of the field energy. The dissipation rate may be written down in terms of the fluctuating rates of strain in the turbulent flow and the fluid's kinematic viscosity, $v$, that has dimensions of action per unit mass. We suggest that the failure to obtain analytical solutions for turbulent processes may solved if these complementary forms of energy are considered.

5.2 *Heat produced by wind farms*

Current practice for wind power makes no provision for heat production other than minimising friction. The radial action theory demonstrates that the back torque exerted by turbines is effectively a work-heat dissipation of wind energy, contribution to its evolution locally at the point of power output. Depending on the factors controlling efficiency, this heat production can be considered as of the same order as the power take-off as electrical energy.

Of more concern could be additional heat release downwind from turbulence. In Table 4, we provided estimates, showing that turbulent release if significantly greater in magnitude. While direct heat production at the turbine is not expected to make a significant difference to air temperature, together with turbulence, a significant fall in the relative humidity of air passing over vegetation and soil together with greater surface interaction by turbulent air can be anticipated. The vortical degree of freedom of motion or action is characterised by its large radius of action, effectively storing latent heat that can be released as radiation in turbulent conditions. It is known that the kinetic energy in laminar flow is not retained in the turbulent motion of air or water moving on much shorter radii of declining scales. Radial action theory predicts this will be the case, the loss of kinetic energy expected as potential energy we have referred to as Clausius' *ergal* [6](Kennedy and Hodzic, 2021b) or internal work is released. The kinetic motion in the system at all scales is sustained by such field or quantum state energy. This is a consequence of the virial theorem, also explained by Clausius.

Impacts of wind farms on surface air temperatures are well documents. Roy and Traiteur [8](2010) claimed that this warming of almost 1 °C compared to an adjacent region resulted from enhanced vertical mixing from turbulence generated by wind turbine rotors. Warmer air from above the surface, particularly at night, was claimed to be brought to the surface. This conclusion was apparently based on an argument that this was the only source of warmer air considered as available. Harris et al. [9](2014) showed "irrefutable night-time warming relative to surrounding areas using observations made from eleven years of MODIS satellite data with pixel size of 1.1 $km^2$. The same conclusion was reached by Miller and Keith [10](2018), showing a significant night-time warming effect at 28 operational US wind farms. They also concluded that wind's warming can exceed avoided warming from reduced carbon emissions for more than a century. According to Miller [11](2020) these effects on warming are detectable tens of kilometres downwind. The opinion that the warming is a result of overturning temperature inverted air at night is not convincing. More direct observations using sensing instruments is required to establish the source of warming.

5.3 *Evidence of vortical entropic potential energy as a heat source released by turbulence*

Chakirov and Vagapov [12](2011) describe a method for direct conversion of wind energy into heat using a Joule machine. They show that turbulence in a rotating fluid with Reynold's number ($R$e) greater than 100,000 provides warmth not obtained when flow is laminar. By insertion of baffles to cause turbulence in the flow path of water set in rotation in a smooth cylinder using direct wind power, they demonstrate that needs for room heating in polar regions can be satisfied. It is well known that the kinetic energy in turbulence is not conserved at lower fractal scales, suggesting that any entropic energy or ergal is also lost in these processes, where work performed is dissipated as heat by friction. We have recently discussed such heat-work-heat cycles in an action revision of the Carnot Cycle [4](Kennedy and Hodzic, 2021a; 2021b), emphasising the importance of entropic field energy as the negative of the Gibbs energy. The molecular kinetic energy in such systems is a small fraction of the total non-sensible heat, stored in quantum state activations of translational, rotation and vibration.

Chervenkov et al. [13](2013) have shown how the kinetic energy and temperature of polar molecules can be reduced with a centrifugal force from around 100 K to 1 K, A redistribution of field entropic potential energy from interior molecules that can be retrieved nearer the centre of the centrifuge is regarded as the cause of the cooling. Geyko and Fisch [14,15](2013) have reported measuring reduced compressibility in a spinning gas where thermal energy is stored in their theory of the piezothermal effect [16](Geyko and Fisch, 2016). This extra heat capacity at constant temperature indicates an additional degree of freedom, that we conclude is vortical, supplementing the well-recognised vibrational, rotational and translational action as degrees of thermodynamic freedom.

The widespread failure to recognise the dominance of this nonsensible field energy as real potential energy (actually, kinetic energy of quanta at light speed [$T = mc^2$, J]) in natural systems, favouring the sensible kinetic heat indicating the temperature of molecules, has been a critical omission [3]. In effect, potential energy in the atmosphere can be gravitational varying vertically, but it can also be stored horizontally. These two forms of energy, kinetic and entropic as negative Gibbs energy, are complementary in operation and we must always have one to have the other. Where viscous dissipation of energy in storms it is not just the turbulence kinetic energy that is released [17] (Businger and Businger, 2001), but the vortical entropic energy much larger in magnitude that sustains the kinetic energy. Of course, it is the current enthalpy sustained by the Gibbs field that actually does the physical damage, as we confirmed for the Carnot cycle [3](2021a).

5.4 *Environmental effects of heat production*

Given the prediction of significant heat production in section 5.1, environmental effects of wind farms should be of concern. In particular, their potential effect on evapotranspiration downwind as a result of turbulence should be considered. Application of the Penman-Monteith equation is the usual method to model evapotranspiration, including evaporation from soil or water surfaces as well as transpiration of water used by plants to absorb nutrients, maintain plant turgor and provide water for photosynthesis. Despite its importance for plant growth in assimilation of carbon dioxide, the actual consumption of water for plant growth is far less than that transpired. The inputs required are daily mean temperature, wind speed, relative humidity and solar radiation. To assist investigation of causes and effects for these events affecting bushfire risk, we are employing the Penman-Monteith equation (*UFlorida* 2020 *AE459*), with data potentially of use from the MODIS satellite.

In this equation $R$n and $G$ indicate solar radiation and local absorption of heat into the soil, $\rho_a$ represents atmospheric density, $C_p$ the heat capacity of air, $e_s^o$ mean saturated vapour pressure (kPa), $r_{av}$ bulk surface aerodynamic resistance for water vapor, $e_s$ mean daily ambient vapor pressure (kPa) and $r_s$ the canopy surface resistance (s m$^{-1}$).

$$ET_{SZ} = \frac{[\Delta(Rn-G)]+[86,400\frac{\rho_a C_p}{r_{av}}(e_s^o - e^a)]}{(\Delta + \gamma(1+\frac{r_s}{r_{av}}))} \quad (15)$$

Wind speed *u* is also included in the numerator. The main drivers of evapotranspiration are heat from solar radiation, plant growth, environmental conditions of temperature and relative humidity as well transport away in air. More important than wind speed, turbulence has now been shown to significantly increase evaporation, as eddy diffusion lengthens the trajectory for water vapor molecules. Since terrestrial wind farms are usually placed in rural areas, we are applying this model to test our prediction that they may contribute to dehydration of the landscape downwind from turbines, increasing fire risk. We will discuss how these proposals may be tested experimentally, including by observations from the MODIS satellite.

To determine the potential effects of wind turbines on evapotranspiration, we calculated evapotranspiration at a range of windspeeds, and then recalculated with a 1 degree C increase in temperature assumed to be the result of wind turbines (Table 6). A 1 degree increase in temperature was used as we found from that turbulence caused in wind at 15 m per sec by wind turbine blades effectively reducing blade vortical speed to 5 m per \sec which could release enough heat to raise the temperature downwind in a swath of 100 m wide and 250 m high more than 50 km downwind by 1 degree Celsius. We used the Penman-Monteith equation for the 5$^{th}$ February at -30.39 latitude, 275 m elevation and assumed an effective daylength of 9.25 hours. The FAO version of the Penman-Monteith equation was applied using the python module ETo, and only the values described here were altered. For both temperature and relative humidity, minimum and maximum values were 20 and 30 C, and 25 and 84%. Evapotranspiration rates calculated without heat input from turbulence induced by the wind turbines are compared with rates corrected with a 1 degree increase in minimum and maximum temperatures, with all else unchanged. The evapotranspiration is between 0.21 mm/day at 5 m/s to 0.58 mm/day at 25 m/s. These predictions, and the short- and long-term effects on dryness of soil and plants, need to be tested experimentally. Clearly, at elevated wind speeds over multiple consecutive warm to hot days, the additional quantity of moisture removed from the soil and vegetation due to wind turbines has the potential to be substantial. Our calculation in Table 6 has taken no account of the downwind turbulence that may increase evapotranspiration significantly [16, 17](Cleugh, 1998; Navaz et al.,2008).

**Table 6.** Predictions regarding heat production and evapotranspiration for 1 $^{o}$C on wind farms

| Wind speed (m sec$^{-1}$) | Evapotranspiration (no wind farm) | Evapotranspiration (with wind farm) | Delta ET (mm day$^{-1}$) |
|---|---|---|---|
| 5.0 | 7.31 mm day-1 | 7.52 | 0.21 |
| 10.0 | 10.18 | 10.52 | 0.34 |
| 15.0 | 12.02 | 12.46 | 0.44 |
| 20.0 | 13.30 | 13.82 | 0.52 |
| 25.0 | 14.24 | 14.82 | 0.58 |

If these results are found in rural landscapes where wind farms are located, they would be of some degree increased risk with respect to optimum agricultural or pastoral productivity. However, in some cases increased temperature and reduced water-holding capacity of air may be beneficial for plant growth, particularly in environments with ample water supplies. In drought prone conditions, negative effects on productivity and increased fire risk can be assumed. There is a need for these factors to be assessed when wind farms are developed.

6. **Conclusion**

Based on a Newtonian approach regarding momentum transfer in the radial action model [20], we have provided an effective method to estimate maximum power extraction from wind turbines. We claim the following advantages for the new theory, all subject to experimental confirmation.

- A more effective mathematical model of wind power output. Using the Carnot approach for power of heat engines, the radial action method allows maximum power to be estimated.
- A better template for wind turbine design is also provided, given the expected close correspondence between theoretical and practical results.

- A means of optimisation of wind power and to minimise heat output is now available [21-23]. This has the potential to be applied as control theory for managing turbines, either solitary or in wind farms.
- Environmental protection is also a possible output from using the action model. Ways to manage turbulence to reduce its possible negative effects can be topics for research.

**Acknowledgements**

**Interests**

**References**

(1) Glauert H. (1935) Airplane propellers. In Aerodynamic Theory. Springer Berlin/Heidelberg Germany.
2. Sorensen JN. (2015) General momentum theory for horizontal axis wind turbines. Springer, Cham Switzerland.
3. Kennedy, I.R., Geering, H., Rose, M. and Crossan A. A simple method to estimate entropy and free energy of atmospheric gases from their action. **2019,** *Entropy* 21, 454-47.
4. Kennedy, I.R.; Hodzic, M. Action and entropy in heat engines: An action revision of the Carnot Cycle. **2021,** 23, 860. https://doi.org//10.3390/e23070860.
5. Kennedy, I.R.; Hodzic, M. Partitioning entropy with action mechanics: Predicting chemical reaction rates and gaseous equilibria of reactions of hydrogen from molecular properties. **2021**, *Entropy*, 23,
6. Schubel, P.J.; Crossley, R.J. Wind turbine blade design. **2012**, *Energies* 5, 3425-3449. doi:10.3390/en5093425.
7. Gupta, S.; Leishman, J.G. Dynamic stall modelling of the S809 aerofoil and comparison with experiments. **2006**, *Wind Energy* 9, 521-547.
8. Roy, S.B.; Traiteur, J.J. Impacts of wind farms on surface air temperatures. Proceedings NAS, **2012,** 107, 17899-17904.
9. Frisch, G. (1995) Turbulence: The legacy of A.N. Kolmogorov. Cambridge University Press.
10. Harris, R.A.; Zhou, L.; Xia, G. Satellite observations of wind farm impacts on nocturnal land surface temperature in Iowa. **2014**, *Remote Sensing* 6, 12234-12246.
11. Miller, L.M.; Keith, D.W. Climatic impacts of wind power. **2018**, *Joule,* 2, 2618-2632.
12. Miller, L. The warmth of wind power. **2020**, *Physics Today* 73, 58-63.
13. Chakirov, R.; Vagapov, Y. Direct conversion of wind energy into heat using Joule machine. **2011**, *International Conference Environmental Computer Science* 19, 12-17.
14. Chervenkov, S.; Wu, X.; Bayerl, J.; Rohlfes, A.; Gantner, T.; Zeppenfield, M.; Rempe, G. Continuous centrifuge decelerator for polar molecules. **2013**, arXiv:1311.7119v1.
15. Geyko, V.I.; Fisch, N.J. Reduced compressibility and an inverse problem for a spinning gas. **2013**, *Physical Review Letters* 110, 150604.
16. Geyko, V.I.; Fisch, N.J. Piezothermal effect in a spinning gas. **2016**, *Physical Review E* 94, 042113.
17. Businger, S.; Businger, JA. Viscous dissipation of turbulence kinetic energy. **2001**, *Journal Atmospheric Sciences* 58, 3793-3796.
18. Cleugh, H.A. Effects of windbreaks on airflow, microclimates and cop yields. **1998**, *Agroforestry Systems* 41, 55-84.

19. Navaz, H,K., Chan, E., Markicevic, B. Convective evaporation model of sessible droplets in a turbulent flow – comparison with wind tunnel data. **2008,** Mechanical Engineering Publications, 149.
20. Kennedy, I.R. *Action in Ecosystems: Biothermodyamics for Sustainability*. Research Studies Press/John Wiley: 2001 Baldock, UK, 251 p., 2001.
21. Kennedy, I.R.; Hodzic, M.; Acharige, N. Designing a regional climate model to test the hypothesis that increasing anthropogenic use of water is a contributor to global warming. **2019**, Australian Meteorological Oceanic Science Conference, Darwin, Australia.
22. Yurdusev, M.A.; Ata, R.; Cetin, N.S.; Assessment of optimum tip speed ratio in wind turbines using artificial neural networks. **2006**, *Energy* 31, 2153-2161.
23. Fitch, A.C. Climatic impacts of large-scale wind farms as parameterized in a global climate model. **2015**, *Journal of Climate*28, 6160-6180.